\documentclass[aip,jcp,reprint]{revtex4-1}
\usepackage{graphicx,dcolumn,bm,amssymb,amsmath}
\usepackage{color,natbib}

\begin{document}

\title{Mechanosensitivity of the II${}^{{\mathbf{nd}}}$ kind: \\ 
TGF-$\bm{\beta}$ mechanism of cell sensing the substrate stiffness}

\author{Max Cockerill, Michelle K. Rigozzi, and Eugene M. Terentjev} 

\affiliation{Cavendish Laboratory, University of Cambridge,  Cambridge, CB3 0HE, U.K.}

%\date{\today}

\begin{abstract}
\noindent Cells can sense forces applied to them, but also the stiffness of their environment. These are two different phenomena, and here we investigate the mechanosensitivity of the II${}^{{\mathrm{nd}}}$ kind: how the cell can measure an elastic modulus at a single point of adhesion -- and how the cell can receive and interpret the chemical signal released from the sensor. Our model uses the example of large latent complex of TGF-$\beta$ as a sensor. Stochastic theory gives the rate of breaking of latent complex, which initiates the signaling feedback loop after the active TGF-$\beta$ release and leads to a change of cell phenotype driven by the $\alpha$-smooth muscle actin. We investigate the dynamic and steady-state behaviours of the model, comparing them with experiments. In particular, we analyse the timescale of approach to the steady state, the stability of the non-linear dynamical system, and how the steady-state concentrations of the key markers vary depending on the elasticity of the substrate. We discover a crossover region for values of substrate elasticity closely corresponding to that of the fibroblast to myofibroblast transition. We suggest that the cell could actively vary the parameters of its dynamic feedback loop to `choose' the position of the transition region and the range of substrate elasticity that it can detect. In this way, the theory offers the unifying mechanism for a variety of phenomena, such as the myofibroblast conversion in fibrosis of wounds and lungs and smooth muscle cell dysfunction in cardiac disease.
\end{abstract}

\maketitle 

%%%%%%%%%%%%%%%%%%%%%%%%%%%%%%%%%%%%%%%%%%%%%%%%%%%%%%%%%%%%%%%%%%%%%%%%%%%%%

\section{Introduction}\label{sec1}

 How cells sense is important to life: homeostasis necessitates sensors.  The relationship between cell morphology and chemical properties of its  environment have long been understood and documented. A plethora of sensory systems embedded in the cell membrane are able to detect specific signaling molecules, giving information to the cell to which it can react. Over the previous two decades there has been growing evidence to suggest that cells use these chemical pathways to sense the mechanical properties of their environment. 

Mechanosensing is a process where at least one chemical reaction in the cell changes in response to a change in its mechanical environment. We specifically wish to distinguish two very different cases:  when external forces are applied to the cell, and when the cell needs to measure the stiffness of its passive environment. We call the former case the `mechanosensitivity of the I${}^\mathrm{st}$ kind'. There is a lot of research on this problem: the cell response to hydrostatic pressure \cite{pressure1, pressure2}, substrate topography \cite{topo} are just some examples; practically, there are implications for biotechnology \cite{guck} and cancer research \cite{cancer1, cancer2}. Ultimately this kind of mechanosensitivity is well understood: there are many molecular processes that respond to an applied force and several different ones are employed by cells in different situations. It has been observed (for example) that mechanotransductive pathways can link compressive forces to translational and post-translational events by cytoskeletal interactions \cite{szafranski2004}, that tyrosine kinases are part of the pathway of the cell response to shear stress \cite{chen1999}, and that mechanical forces can alter ion channel permeability \cite{pivetti2003}. In each of these cases we find a remarkable protein design to transduce a mechanical stimulus into a chemical pathway that the cell can understand.

It is much more challenging to understand the mechanism of `mechanosensitivity of the II${}^\mathrm{nd}$ kind', when the cell responds to substrate stiffness \cite{yeungGeorgesJanmey_2005, stem}. 
In order to detect a linear response function (elastic constant or viscosity), the cell has to apply a force of its own, and two separate measurements have to be carried out: of this force or stress as it is transmitted to the substrate; and of the displacement or strain in the deforming substrate. The ``stiffness'' is  a ratio relating these two physical parameters. For this kind of measurement, the apparatus requires two points of application: to measure displacement a fixed reference point is always needed -- to measure force one has to ensure a point of reaction as well. We need \textit{two} fingers to squeeze a test object from two sides to determine how stiff it is -- in the same way, all engineered devices ultimately have two points of action on the test sample. A mechanical sensor like the latent transforming growth factor-$\beta$ complex we study in this paper (lc-TGF-$\beta$) is effectively a single localised molecular protuberance outside the cell surface; it only has one point of application through which a pulling force is transmitted from the cell interior, and no {\it a priori} information about the reference point in the substrate \cite{shi2011}. An attempt to use a dynamical regime to overcome the need for a separate reference point in space also does not work: it is widely known that cells exert a constant force on their environment, not a dynamically varying force  \cite{demboWang1999,schwarzBalaban2002,tanTienChen2003}. A recent paper by Escud\'{e} \textit{et al} \cite{escude_rigozzi_terentjev2014} have formulated a new model, using the viscoelastic  response of the substrate coupled with the stochastic process of rupturing the lc-TGF-$\beta$ by a constant pulling force generated by the cell. The result of this model is the predicted rate of free TGF-$\beta$ release, expressed as a compact function of the pulling force and the substrate stiffness. 

Cells in the body and in-vitro are normally supported biochemically and mechanically
by the extracellular matrix (ECM), or its fragments (such as fibronectin) bound to artificial substrates. Experiments have observed that the stiffness of a substrate can stimulate differentiation of a fibroblast into a myofibroblast, or determine the future lineage of stem cells \cite{englesSen2006}: whether they will differentiate to bone, muscle, or nerve cells for example.  How does the cell use proteins to sense stiffness and how does it convert a measurement it might be able to make into a chemical signal that the cell can understand and respond to? And what might a robust system which is designed to answer these two questions look like?

In this work we develop a coupled set of kinetic equations that describe a dynamical signalling feedback loop capable of interpreting the measurement of the substrate stiffness and convert it into the morphological changes the cell needs in reaction to such a substrate. We follow the earlier work by using the lc-TGF-$\beta$ family as a particular biological sensor, shown in \cite{wipffRifkinHinz_2007} to be associated with the fibroblast to myofibroblast differentiation, and in \cite{sinhaHeagertyKielty_2002,sinhaHoofnagleOwens_2004} with the development of smooth muscle cells from embryonic stem cells, as well as their transformation following arterial injury or significant deposition of mechanically rigid plaque. We combine the result on the stiffness-dependent rate of latent complex rupture with the analysis of the coupled  biological system to derive a set of  equations to describe the dynamics of the protein pathway which forms the mechanosensing signalling loop. We compare the results and predictions to the stiffness-dependent behaviour observed in cell-adhesion experiments, as well as in-vivo observations, and find a strong correlation of the model behaviour and real systems.

In the discussion of the biological system and comparison to experimental results we will appeal most significantly to the fibroblast system for guidance. This is because a substantial volume of research has been carried out into the behaviour of this system and because the model developed in \cite{escude_rigozzi_terentjev2014}, which we incorporate into our model, bases its arguments on the fibroblast system too. Besides, fibroblasts are the most prevalent cells in connective tissue and are vital in tissue repair within the body. When a break in the tissue occurs fibroblasts differentiate to myofibroblasts which act to pull the wound back together. The highly contractile myofibroblast form is produced by high expression of $\alpha$-smooth muscle actin ($\alpha$-SMA) fibres \cite{TomasekGabbianiHinzChaponnierBrown2002}, the production of which has been shown to be induced by TGF-$\beta$ \cite{wipffRifkinHinz_2007,hinz2001alpha}.

\begin{figure}
\centering{
\includegraphics[width=0.7\columnwidth]{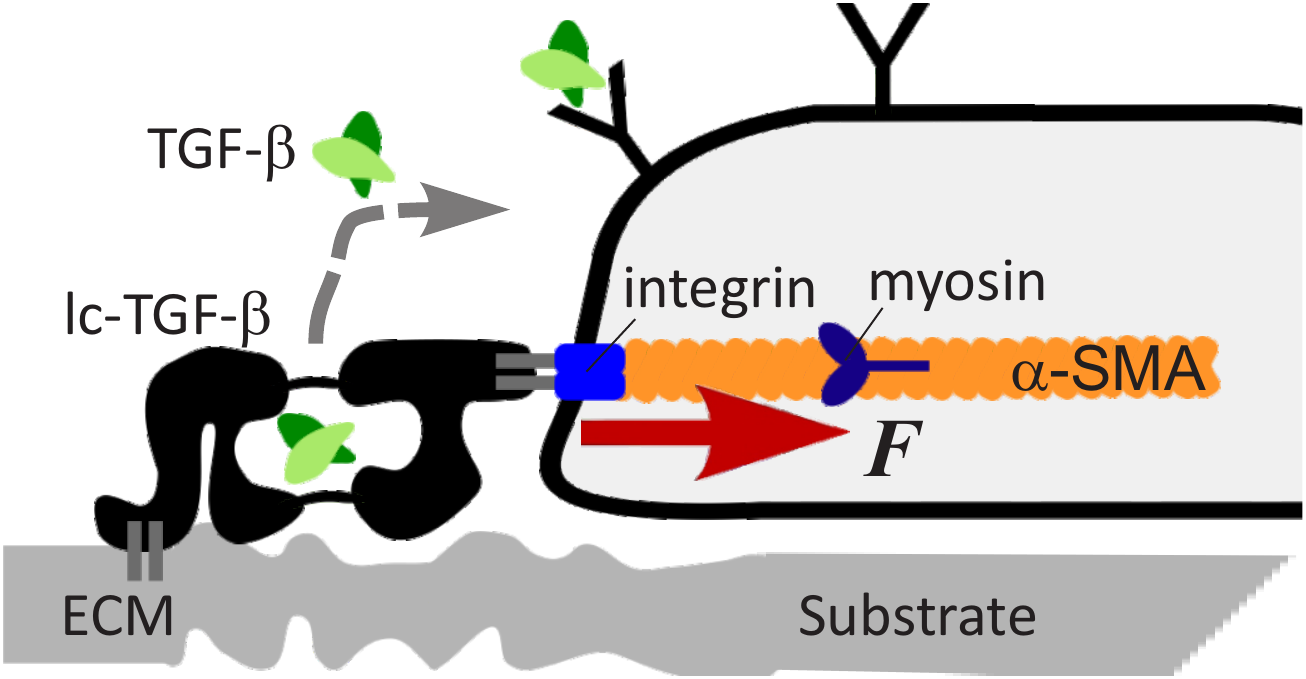}
%\rule{2cm}{2cm} % added by Karol Kozioł
}
\caption{A diagram showing the mechanosensor system, from \cite{escude_rigozzi_terentjev2014}. The large latent complex of TGF-$\beta$ binds to ECM proteins attached to a substrate, which may be deformable. The other end of lc-TGF-$\beta$ is attached to the cytoskeleton via an integrin complex that transmits the pulling force. If released, the free active TGF-$\beta$ can bind to a set of receptors on the cell surface to initiate the signalling loop discussed in this paper. \label{Diagram: Mechanosensing Model_Biological}}
\end{figure}

The mechanosensing system is summarised diagrammatically in figure \ref{Diagram: Mechanosensing Model_Biological}. The latent complex of TGF-$\beta$ is the stretch-dependent mechanosensor; it  is representative of a larger family of mechanosensor complexes. There are three isoforms of the signalling molecule: TGF-$\beta1$, 2 and 3 with distinct effects in vivo, but their complexes are all produced in the same way: tightly but noncovalently bound as a dimer to their  latency associated peptide (LAP-1,2,3) to form the `small latent complex'. This forms a larger complex (which we are referring to as  lc-TGF-$\beta$) by disulphide links between certain domains in LAP and the latent TGF-$\beta$ binding protein (LTBP), which in turn can bind to the ECM to anchor the complex on the substrate \cite{roberts2001}.  The  large latent complex as a whole is secreted out of the cell, where it may connect LTBP to ECM components. Though no unambiguous binding partner for the complex has been found in the ECM \cite{Annes03}, there is evidence in favour of strong covalent binding, likely to 8-Cys domains on LTBP \cite{Unsold01} and it has been known to bind to certain ECM proteins such as fibrillin and fibronectin \cite{BuscemiRamonetKlingberg2011}. We shall assume that this bond is strong and not breaking during the processes we study.

%For inserting figures, all the images are in current working folder and insert the below command
%for calling figures. We have to insert the image wherever the first reference in the text.

 On the other side, the lc-TGF-$\beta$ adheres to integrins on the cell surface. which are bound to the actin fibres making up the contractile cytoskeleton, allowing the pulling force generated by myosin motors to be transmitted to the complex, and through it -- to the substrate. Upon mechanical rupturing of the complex, the active TGF-$\beta$ is released. Once the lc-TGF-$\beta$ is spent, it can no longer function as the adhesion point and may or may not be replaced by a new latent complex.

Once TGF-$\beta$ is released, it is free to diffuse. It could therefore bind to specific receptors on the cell surface. Binding initiates the signalling loop in the cell illustrated in figure \ref{Diagram: Protein Interaction} below. It may stimulate production of new lc-TGF-$\beta$ units to replace those that are spent, and also add more $\alpha$-SMA on the inside of the cell \cite{wipffRifkinHinz_2007}. $\alpha$-SMA is produced to form more actin filaments and thus increase the pulling force $F$ applied to the latent complex adhesion points. We shall assume a simple linear relationship between the concentration of $\alpha$-SMA and the force $F$ applied to lc-TGF-$\beta$.

\section{Mechanosensing reaction kinetics}

\subsection*{Rate of latent complex breaking}
  To describe the underlying physics of the system we start with the rupturing of lc-TGF-$\beta$ by an external pulling force $F$. On the far side this complex is attached to a viscoelastic substrate which we model by a simple Voigt model with an effective stiffness $\kappa$, that is directly related to the Young modulus $E$ of the elastic medium via the classical continuum-mechanics relation: $\kappa \approx \frac{4}{3}\pi E \xi$ \cite{landauLifshitz_1986}. Here $\xi$ is the characteristic length scale of short-distance cutoff in the Green's function of linear elasticity. We shall find that our results best match experimental observations when this length scale is taken as $ \xi \approx100\mathrm{nm}$, which happens to be a characteristic mesh size of the extracellular matrix of crosslinked filaments. Then a rigid glass substrate, with a Young modulus $E \approx 50$\,GPa, gives the maximum value of our $\kappa \approx 21,000\, \mathrm{N/m}^2$. A typical muscle, organ or connective tissue (as well as synthetic gels) have $E \approx 10-50$\,kPa, giving the range of $\kappa \approx 0.004-0.02\, \mathrm{N/m}^2$. At the other end of rigidity scale, the brain tissue is very soft: $E \approx 100$\,Pa, leading to the lower bound of  $\kappa \approx 4 \cdot 10^{-5} \mathrm{N/m}^2$.

Overdamped thermal fluctuations dominate the response of this series of mechanical elements. In equilibrium (i.e. with no force applied) the probability (rate) of spontaneous breaking of the latent complex is proportional to $\exp [-\Delta G/ k_BT]$ with $\Delta G$ the bonding free energy of the large latent complex and $k_BT$ the thermal energy of the reservoir. When the pulling force is applied to the complex, it is also transmitted to the deformable substrate and the effective potential holding the latent complex together changes.

The rate of breaking of the latent complex attached to a viscoelastic substrate, under a constant pulling force, is given by the expression below (which corrects a typo in the original work \cite{escude_rigozzi_terentjev2014}):
\begin{equation} \label{eq:Kmech0}
k_{m}(F, \kappa) = C \frac{\tilde{\kappa}^{{1}/{2}} \Delta_F^{-1/4} }
			 { \left( \Delta_F^{{1}/{4}} + (\frac{1}{6}\tilde{\kappa})^{{1}/{2}} \right)}
			 \frac{ e^{- F^2/2 \kappa k_BT} }{ 1+2e^{{\Delta_F^{3/2}\Delta G}/{k_{B}T}} }   , 
\end{equation}
where the shorthand non-dimensional notations used are:
\begin{eqnarray}
C = \frac{3 \tilde{D}}{u_{0}^2} \left(\frac{\Delta G }{2\pi \, k_B T}\right)^{1/2} , \ \ 
\tilde{\kappa} = \frac{\kappa u_0^2}{\Delta G} \,\,\,, \ \ \Delta_F 	= 1-\frac{2Fu_{0}}{3\Delta G}. \nonumber
\end{eqnarray}
Here  $u_0$ is the characteristic distance the latent complex needs to be stretched for rupturing. We could estimate the bond free energy $\Delta G$ holding the large latent complex together as for a few hydrogen bonds \cite{hummberSzabo_2003}, $\Delta G \approx $ 10-16\,kcal/mol, and the stretching distance before rupturing as the size of an aminoacid residue, $u_0 \approx 0.3$\,nm. This is consistent with descriptions of force induced TGF-$\beta$ activation by binding of $\alpha_v\beta_6$ to the latent complex \cite{shi2011}. In this case the range of the scaled (non-dimensional) $\tilde{\kappa}$ is between $4\cdot 10^4$ at the upper bound of rigid glass, $0.02$ for a typical gel or muscle tissue, down to $8\cdot 10^{-5}$ at the lowest bound of $E=100$\,Pa.  

The effective diffusion constant $\tilde{D}$, which enters as a common factor in the expression for the rate of mechanical breaking $k_m$, is the ratio $k_BT/\sqrt{\gamma_1 \gamma_2}$ reflecting the geometric mean of the damping coefficients (loss factors) of the substrate and the latent complex (see \cite{escude_rigozzi_terentjev2014} for detail). We will find it necessary to estimate the value of this effective diffusion constant, which is determined by the effective damping coefficient of the viscoelastic substrate medium $\gamma \approx \kappa \tau_\mathrm{R}$ (that is, the characteristic time of stress relaxation in the medium). There is a broad range of stress relaxation times found in different material systems, but for the sake of simplicity we shall take the characteristic Rouse time for a polymer network: $\tau_\mathrm{R} \approx 10^{-5}$s; whether it is a synthetic polymer gel or an organic tissue -- this may serve as a first estimate (we must be aware that such a simple argument is likely to fail in the limit of very rigid glass and also in very soft gels/tissues with extra-long polymer strands). Then for a mid-range substrate stiffness of a typical gel or muscle tissue ($E=20$\,kPa), at room temperature we will have $\tilde{D} \approx 5 \cdot 10^{-14}\mathrm{m}^2/\mathrm{s}$, a sensible value lower than the typical constant of thermal diffusion in water.  We take the form of $\tilde{D}$ to be $k_BT/\kappa \tau_\mathrm{R}$ in the numerical analysis below.

\begin{figure}
\centering{\includegraphics[width=\columnwidth]{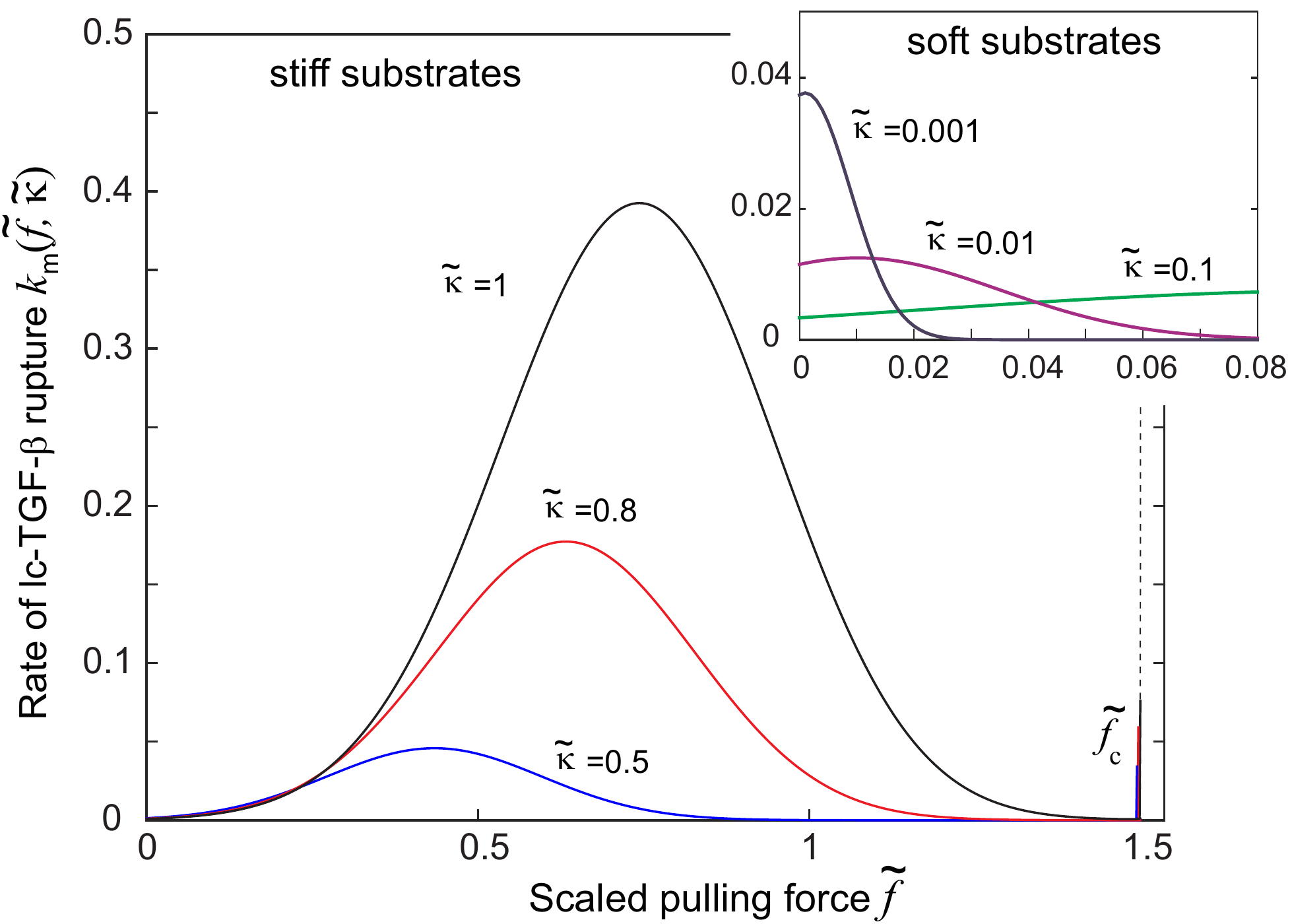}}
\caption{Rate of rupture of lc-TGF-$\beta$ (in real units of [s${}^{-1}$]) plotted against the non-dimensionless scaled force $\tilde{f} ={Fu_{0}}/{\Delta G}$, at various different values of dimensionless elasticity $\tilde{\kappa}$. The plot highlights the dramatic increase of the rate on stiffer substrates, and also the highly non-monotonic variation with force. The inset shows that on very soft substrates the rate of rupture is very low and in fact decreases further with the force.  Note that the rate always diverges at $\tilde{f}_c=3/2$, which is the maximum force that the latent complex can stand.  \label{plot:Rate_Vs_f}}
\end{figure}

In spite of its apparent complexity, the expression for the rate of mechanical breaking of the latent complex has only a few straightforward features: it describes the latent complex as having the physical bond energy $\Delta G$ in equilibrium, which is decreased when a pulling force is applied to it, making the basis for the Bell formula for the rate \cite{Bell}: when the energy barrier is large, the applied force is small, and the substrate stiffness is low -- eq. \eqref{eq:Kmech0} does reduce to:
\begin{equation} \label{eq:KmechBell}
k_{m}(F, \kappa) \approx \frac{1}{2} C\, \exp \left[ -  \frac{\Delta G - F u_0}{k_BT} \right].
\end{equation}
However, we cannot use this Bell formula since we need to consider high pulling forces and moderate barriers that occur in practice.

On the other hand, eq. \eqref{eq:Kmech0} also contains the effect of the deformed substrate, which is also subject to the same force $F$, transmitted through the latent complex while intact. This produces the `enzyme effect' discussed in  \cite{escude_rigozzi_terentjev2014}: on increasing the pulling force and deforming the substrate into a new (deeper) elastic  energy miniumum, the fluctuations of the point of adhesion become confined; this is expressed by the factor $\exp [ - F^2/2 \kappa k_BT]$ in eq. \eqref{eq:Kmech0}. The other effect of the substrate viscoelastic response is to alter the non-exponential prefactor of the rate expression by adjusting the statistical weight of the intact latent complex in the full ensemble. The dependence of the breaking rate, $k_m$, on the applied force for several values of substrate stiffness is illustrated in figure \ref{plot:Rate_Vs_f}.

At sufficiently high substrate stiffness, expressed by the non-dimensional scaled parameter $\tilde{\kappa}$ given below the eq. \eqref{eq:Kmech0}, there is a maximum in the rate of breaking of the latent complex.  So, initially, the positive feedback is expected -- an increase in the pulling force increases the rate of latent complex breaking, leading to a further increase in force, etc. until the cell finds the equilibrium. However, at very low values of $\tilde{\kappa}$ the maximum is at zero applied force. That is, on such soft substrates a pulling force actually reduces the rate of latent complex breaking causing the negative feedback. The outcome depends on the details of interactions between the key parameters in this feedback loop.

\subsection*{Protein interactions}
Upon rupturing of one latent complex, an active TGF-$\beta$ unit is released and becomes free to diffuse towards a specific receptor type on the cell membrane, and bind to it. The binding of TGF-$\beta$ stimulates production of $\alpha$-SMA \cite{AroraNaraniMcCulloch} by a signalling pathway that we do not go into details of, except to know the ratio of how many $\alpha$-SMA monomers are produced to each TGF-$\beta$ bound. Given the linear relationship between $\alpha$-SMA and force $F$ applied to lc-TGF-$\beta$, this stimulated production results in an increased force on the latent complex. The increase in applied force changes the rate of rupture of the complex, with the feedback nature depending on the regime of $k_m (F,\kappa)$.

\begin{figure}
\centering{\includegraphics[width=0.9\columnwidth]{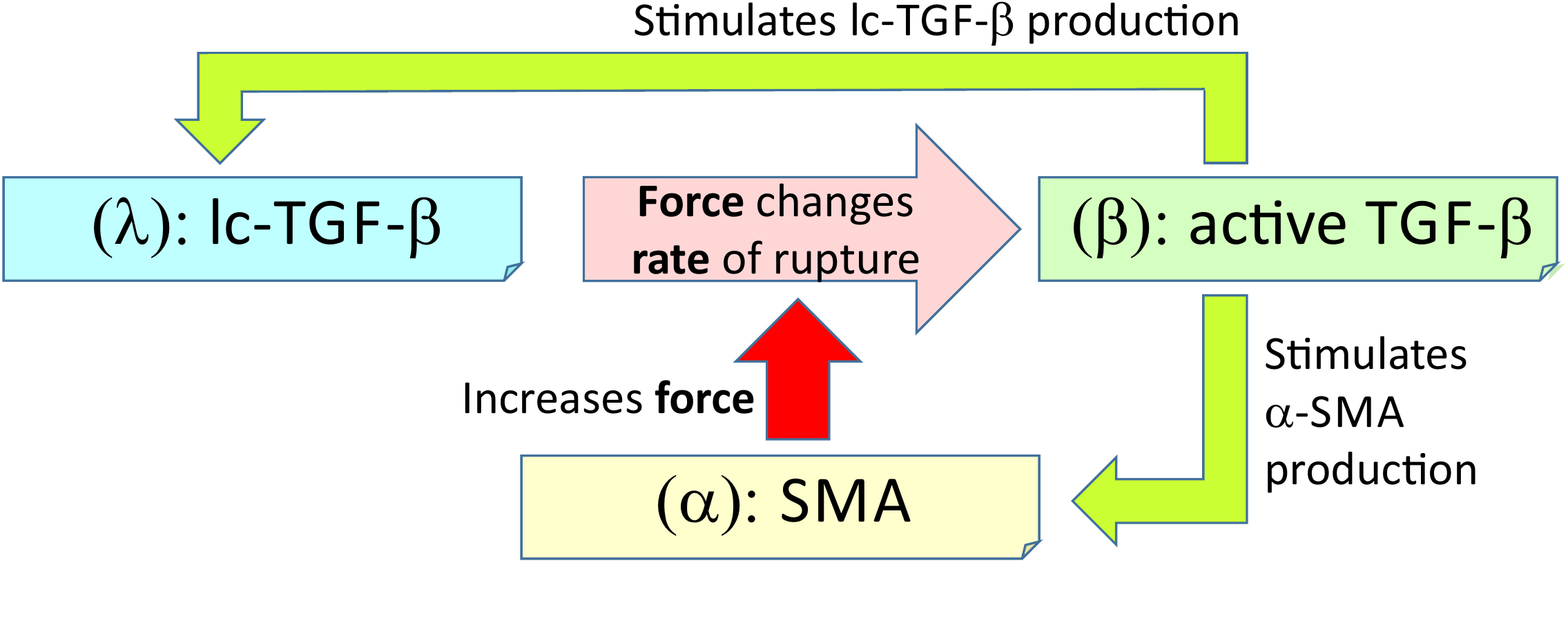}}
\caption{A diagram showing how the proteins in our mechanosensing model interact. The three key markers of the signalling loop are: $\lambda$ (the concentration of lc-TGF-$\beta$), $\beta$ (the concentration of active TGF-$\beta$) and $\alpha$ (the concentration of SMA). The rate of spontaneous breaking of $\lambda$, $k_m(\alpha, \kappa)$, is a non-linear function of the pulling force $F$ applied by the cell, eq. \eqref{eq:Kmech0}.  \label{Diagram: Protein Interaction}}
\end{figure}

 We now seek to mathematically model the system of protein interactions described above. The three key proteins are $\alpha$-SMA (labelled $\alpha$), TGF-$\beta$ (labelled $\beta$) and lc-TGF-$\beta$ (labelled $\lambda$), see figure \ref{Diagram: Protein Interaction}. To be specific with dimensionality, we consider the evolution of the number per cell of each of these proteins, $\alpha(t), \ \beta(t)$ and $\lambda(t)$. The number per cell can be re-dimensionalised into a concentration by considering the typical mass of the protein in question and the volume of the cell -- e.g. a typical fibroblast.

Within our model there are certain assumptions about the state of proteins that we count in our variables. $\lambda (t)$ represents only those lc-TGF-$\beta$ that are attached to both the cell and the substrate and have not ruptured. $\beta (t)$ includes only TGF-$\beta$ molecules that are released from their latent complex and are close to the cell surface, but not yet bound to a receptor.  We can describe the evolution of these three variables in the following set of differential equations:
\begin{eqnarray}  \label{eq:differential equations without sinks}
	\frac{d\lambda}{dt} &=& 	-k_{m}(\alpha,\kappa)\lambda 	+ A_{\lambda}k_{\beta}\beta  \label{eq:differential equations without sinks1} \\
	\frac{d\beta}{dt} 	&=& 	+k_{m}(\alpha,\kappa)\lambda 	- k_{\beta}\beta  \label{eq:differential equations without sinks2} \\
	\frac{d\alpha}{dt} 	&=& 	+A_{\alpha}k_{\beta}\beta  \label{eq:differential equations without sinks3} .
\end{eqnarray}
Here the coefficients $A_{\lambda}$ and $A_{\alpha}$ are, respectively, the number of fresh lc-TGF-$\beta$ complexes and $\alpha$-SMA monomers produced for every instance of TGF-$\beta$ binding to a receptor. Implicit within these two quantities are all the underlying protein pathways whose starting point is a TGF-$\beta$ binding to its receptor, and whose end point is the production of a new latent complex and $\alpha$-SMA subunit, respectively. $k_{\beta}$ is the rate of binding of TGF-$\beta$ to the cell receptors. The key element of this model is the rate of the rupture of lc-TGF-$\beta$, given in eq. \eqref{eq:Kmech0} and  discussed above.

We have to assume a linear relationship between $\alpha (t)$ and the pulling force $F$  to be of the form:
			\begin{equation}\label{eq:Force alpha relationship}
					F=\zeta f_m n_m \alpha  ,
			\end{equation}
where $\zeta$ is the fraction of  $\alpha$-SMA  monomers polymerised into actin filaments that are involved in the application of force on $\lambda$ (this is expected to be a small number). $f_m$ is the force exerted by a myosin motor, which is a known parameter for each individual motor  ($\sim 2$\,pN), but may be a function of ATP availability in the vicinity of the adhesion complex;  $n_m$ is the number of myosin motors that act along each filament on average, a parameter that the cell can adjust within a wide range.

\subsection*{The dissipative sinks}
		We must also consider certain dissipative terms that affect all three of our dynamic variables, so that the system can find a homeostatic state; three dissipative terms are relevant, one for each of the three proteins. 

		The $\lambda$ sink is the saturation of integrin binding sites. We expect there to be only a certain number of sites over the surface of cell in contact with the substrate, at which the latent complex can attach to both the cytoskeleton and the substrate. This number may be large, but it is nevertheless limited, and this restricts the maximum value that $\lambda$ can take. We consider a probability of a lc-TGF-$\beta$ attaching to the substrate and cytoskeleton which has the form:
			\begin{equation}
					P(\lambda,N)=P_0\frac{N-\lambda}{N} ,
			\end{equation}
where $N$ is the total number of available binding sites on the cell surface and $P_0$ is the probability of a latent complex attaching to the substrate and cytoskeleton when $N=\infty$ (i.e. there is no limitation on $\lambda$). This probability should be multiplied by the rate of production of lc-TGF-$\beta$ ($A_{\lambda}k_{\beta}\beta$) to give a total rate of production of $\lambda$, to replace the second term in eq. \eqref{eq:differential equations without sinks1}:
			\begin{equation}\label{eq:Lambda rate of production}
					\frac{d\lambda}{dt} =  A_{\lambda}k_{\beta}P_0\frac{N-\lambda}{N}\beta . 
			\end{equation}

Given the context of the problem, we expect the free TGF-$\beta$ molecules to not only diffuse towards the cell receptors but also could drift away from the cell entirely. Therefore, a $\beta$ sink due to diffusion away from the cell surface should be included. We consider this diffusion loss by treating TFG-$\beta$ as a free Brownian particle diffusing away from a sphere (representing the cell surface). By considering the Gaussian probability for a molecule to be at a position $r$ from the surface after time $t$, then setting $r=0$ we find the concentration near the surface to decrease as $P(r=0)=\ell(2\pi Dt)^{-1/2}$,  where $\ell$ is some normalisation constant with units of length. 

 Our variable $\beta (t)$ is then considered to be the probability of finding the TGF-$\beta$ particle at $r=0$. Initially we expect TGF-$\beta$ to be present only at the cell surface, where it was released from the latent compolex. To normalise to unity at $t=0$ we need to identify the radial lengthscale on which TGF-$\beta$ exists. Because the probability tends to a delta function as $t$ tends to 0, we assume that the radial lengthscale for normalisation is the diameter of TGF-$\beta$ molecule: $\ell=2R_{\beta}$.

 Considering the probability as a volume fraction, we can write $P(0,t=0)=\beta V_{\beta}/V_\mathrm{tot}$ where $V_{\beta}$ is the volume of one TGF-$\beta$ subunit and $V_\mathrm{tot}$ is the total volume occupied by all $\beta$ at $t=0$: a spherical shell at the radius of the cell ($R_\mathrm{cell}$) of thickness $2R_{\beta}$. Therefore we want the contribution to the differential equation \eqref{eq:differential equations without sinks2} to yield: $\beta (t) =2R_{\beta}V_\mathrm{tot}/(V_{\beta}\sqrt{2\pi Dt})$, when no other terms are present. Differentiating  gives a rate of diffusive loss: 
			\begin{equation}\label{eq:Beta diffusion term}
					\frac{d\beta}{dt}=-\frac{\pi DV_{\beta}^2}{4R_{\beta}^2V_\mathrm{tot}^2} \beta^3 .
			\end{equation}		
For our calculations it will be important to estimate the coefficient in this relation, which involves several known length scales; the values for $D$, $R_\mathrm{cell}$ and $R_\beta$ are given in table below. 

In the cell we always find that $\alpha$-SMA decays over time, being used in various other processes. The rate of this spontaneous decay has been measured experimentally and also listed in the table below; we denote it as $k_{\alpha}$:
			\begin{equation}\label{eq:Alpha decay term}
					\frac{d\alpha}{dt}=-k_{\alpha}\alpha . 
			\end{equation}
			
Combining all of the above together we arrive at the final set of three rate equations, which are now have balanced sources and sinks for the evolution of all three of our markers:

\begin{eqnarray}
\frac{d\lambda}{dt}	&=& 	-k_{m}(\alpha,\kappa)\lambda
						+ A_{\lambda}P_0 k_{\beta}\frac{N-\lambda}{N}\beta 	\label{eq:lambda differential equation with sink} \\
\frac{d\beta}{dt}    &= &	+k_{m}(\alpha,\kappa)\lambda - k_{\beta}\beta	-\frac{\pi DV_{\beta}^2}{4R_{beta}^2V_{tot}^2} \beta^3 	\label{eq:beta differential equation with sink}\\
\frac{d\alpha}{dt} 	&=& 	+A_{\alpha}k_{\beta}\beta 	- k_{\alpha}\alpha
				\label{eq:alpha differential equation with sink}
\end{eqnarray}

{	\subsection*{Fixing parameter values} \label{Fixing parameter values}

 \hspace{10pt}The non-linear kinetic model of the signaling feedback loop has many parameters. Where available, values of experimentally derived parameters were always used in preference.  The table below gives the values of parameters that indeed have been determined experimentally.

%The Table \ref{tab1} how to code the tables and place the table where the first
%table reference in the text. See below example table coding

\begin{table}[h]
\begin{center}
\begin{tabular}{@{\vrule height 10.5pt depth3.8pt  width0pt}ccc}
\multicolumn{3}{l}{{\bf \hspace{-8.5pt} Experimentally determined values}} \\
\hline
\vrule depth 6pt width 0pt \bf Parameter & \bf Value & \bf Origin\\
\hline
$R_{\rm cell}$			&	$5 \, \rm \mu m$					& \rm \cite{Freitas1999}\\
$k_{\alpha}$		&	$6\cdot10^{-6}\rm s^{-1}$				& \rm \cite{AroraNaraniMcCulloch} \\
$f_m$ 				&  	$2.3\cdot10^{-12}$\, $\rm N$ 		& \rm \cite{YanagidaIshijima1995}\\
$n_m$ 				& 	$80$ 					& \rm \cite{Cooper2000}\\
$N$ 				& 	$1000$ 			& \rm \cite{BenedettoPulitoCrichTaroneAimeSilengoHamm2006}\\
$D$ 				& 	$7.7\cdot10^{-12}\, \rm m^2s^{-1}$ 		& \rm \cite{PrendergastMann1978,Khalil1999}  \\
$R_{\beta}$			&	$1.66 \, \rm nm$				& \rm  \cite{hink2000} \\
\hline
\end{tabular}
\end{center}
\end{table}

\noindent The value of $k_{\alpha}$ was obtained from the line in figure 5 of \cite{AroraNaraniMcCulloch} representing decay of $\alpha$-SMA without TGF-$\beta$ in a cell anchored to a gel. By assuming exponential decay between day 1 and day 5, a value was approximated which fits this curve. 
The value of $D$ is the diffusion constant for Green Fluorescent Protein (GFP), which is comparable in mass to TGF-$\beta$, and so it is expected that the values of both diffusion parameters are very close. 
No value of $R_{\beta}$ could not be found in the literature so it was approximated by assuming that GFP (a cylindrical protein) and TGF-$\beta$ have similar volumes and that TGF-$\beta$ is spherical. The diameter and length of GFP are $2.4\rm nm$ and $4.2\rm nm$ respectively.  The values of $\tilde{D}$, $\Delta G$ and  $u_0$ for the large latent complex mechanosensor itself have been discussed earlier in the text. 

Other parameters were not experimentally defined, so values that seemed physically reasonable were chosen and the dependence of system behaviour on these values checked.

\begin{table}[h]
\begin{center}
\begin{tabular}{@{\vrule height 10.5pt depth3.8pt  width0pt}cc}
		\multicolumn{2}{l}{{\bf \hspace{-8.5pt}  Model assumptions}} \\
		\hline
		\vrule depth 6pt width 0pt \bf Parameter & \bf Value \\
		\hline
$P_0A_{\lambda}$	&	$1.7$			 \\
$A_{\alpha}$		&	$3$				 \\
$\zeta$				&	$1\cdot10^{-4}$		 \\
		\hline
\end{tabular}
\end{center}
\end{table}

The value for $P_0A_{\lambda}$ was selected by assuming each TGF-$\beta$ binding to a receptor stimulated production of approximately 1 or 2 new lc-TGF-$\beta$. We expect all (or almost all) the lc-TGF-$\beta$ produced to be used in this mechanosensing system. We also assume that $P_0$ is close to 1. A value of 1.7 was selected to approximately reflect these assumptions.

$A_{\alpha}$ was selected to roughly reflect the high concentration of $\alpha$-SMA in the cell. In order for the cell to react to a stimulus it will need to produce a significant amount of $\alpha$-SMA monomers.

The value of $\zeta$, the fraction of newly produced $\alpha$-SMA that ends up polymerised into F-actin and contribute to the pulling force on the adhesion complex,  was chosen by considering the ratio of the volume of an $\alpha$-SMA filament monomer (modelled as a sphere) to that of an $\alpha$-SMA filament subunit (modelled as a cylinder). We define the length of a subunit as the step size of myosin which is $8$\,nm \cite{toyoshima1990}. The diameter of actin filaments is about $8$\,nm \cite{Cooper2000}. The average diameter of an $\alpha$-SMA monomer is about $4.8$\,nm \cite{kabsch1990}. This gives a ratio of $0.14$. We then multiply this ratio by the fraction of $\alpha$-SMA being used in this process to total $\alpha$-SMA in the cell. This was estimated roughly to be $1/1500$, but whatever this value -- it is reassuring that this number is small since one does expect that only a small fraction of new  $\alpha$-SMA monomers would contribute to the increase of the pulling force on the point of adhesion.

The most difficult parameter for us to estimate is the rate of binding of free TGF-$\beta$ to its receptors on the outer cell surface, $k_{\beta}$. It is not experimentally determined and is difficult to define through arguments of plausibility. We fixed its value by requiring that the transition observed in steady-state values of $\lambda$, $\beta$ and $\alpha$ (discussed in the following sections) corresponds to experimental results, so in a way $k_{\beta}$ is the only true fitting parameter of our model.  In culture the fibroblast transition occurs at a Young modulus value $E\approx 11$ kPa \cite{wipffRifkinHinz_2007}. Using $\kappa=\frac{4}{3}\pi E\xi$  (as above, with $\xi\approx 100 \rm\,nm$) we find the corresponding dimensionless (scaled) elastic modulus: $\rm \tilde{\kappa}_{transition}\approx 5.9\cdot 10^{-3} $. To reproduce this value in our results, the rate $k_{\beta}$ needs to be in the range:

\begin{table}[h]
\begin{center}
\begin{tabular}{@{\vrule height 10.5pt depth3.8pt  width0pt}cc}
		\multicolumn{2}{l}{{\bf \hspace{-8.5pt} Constrained parameters}}\\
		\hline
		\vrule depth 6pt width 0pt \bf Parameter & \bf Value \\
		\hline
$k_{\beta}$			& 	$\hspace{14pt}5\cdot10^{-5}\,s^{-1}\hspace{14pt}$ 	\\
		\hline		
\end{tabular}
\end{center}
\end{table}

\subsection*{Method of solving. }
All studies of the differential equations were done in Matlab using the `ode23s' ordinary differential equations numerical solver package provided. Initial conditions were always set randomly, unless otherwise stated. For a given set of initial conditions the system always relaxed to the same steady state values. For $\tilde{\kappa} \geq 1$ the solver would often take a very long time to produce a result, depending on initial conditions, because of the extreme stiffness of the system. Combinations of initial conditions and $\tilde{\kappa}$ that took a long time to solve were avoided for the majority of the calculation. However, the behaviour was checked with one or two `awkward' combinations to ensure that it agreed with other results and trends given in this paper. Note that to ensure $\lambda$ reached steady state, a very long time span had to be used.

\section{Evolution of variables}

The variables $\alpha, \beta$ and $\lambda$ evolve in time from various initial conditions to the final steady state.  Evolution maps of the variables are shown in figure \ref{plot:Variable Evolution Map (3D)} for low $\tilde{\kappa}$ and a reasonably high $\tilde{\kappa}$. The value for which we consider $\tilde{\kappa}$ to be `high' or `low' will be discussed later.  It is clear that the dynamical system is robust and has a single stable attractor for a wide range of parameters (we have tested a large number of parameter values, not just the ones given in the tables above). The position of this fixed point attractor is dependent on $\tilde{\kappa}$, that is, a single fixed point for each value of $\tilde{\kappa}$, and therefore provides a vehicle for mechanosensing of the II${}^\mathrm{nd}$ kind.

\begin{figure}%[h]
\centering{
\includegraphics[width=0.9\columnwidth]{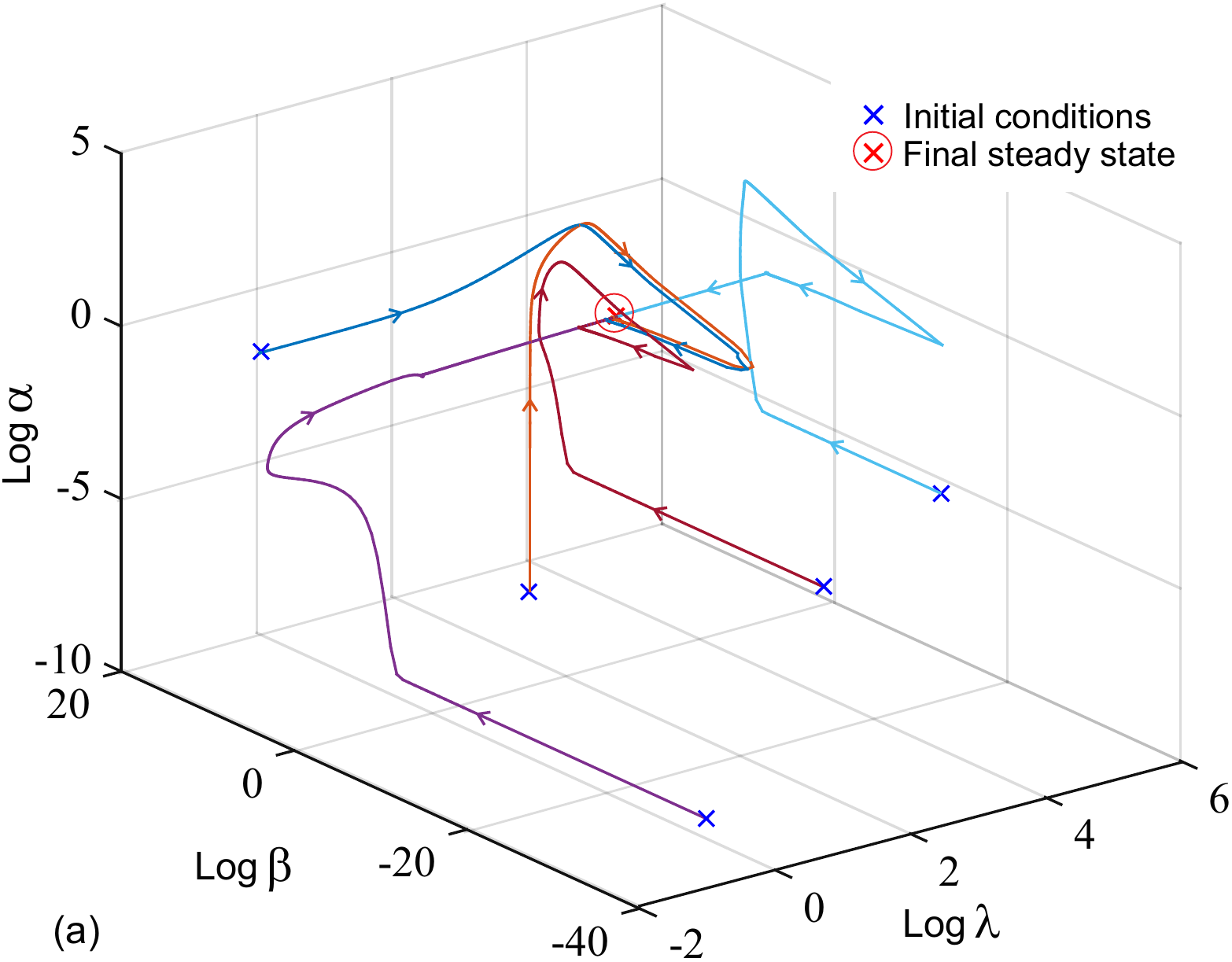}
\includegraphics[width=0.9\columnwidth]{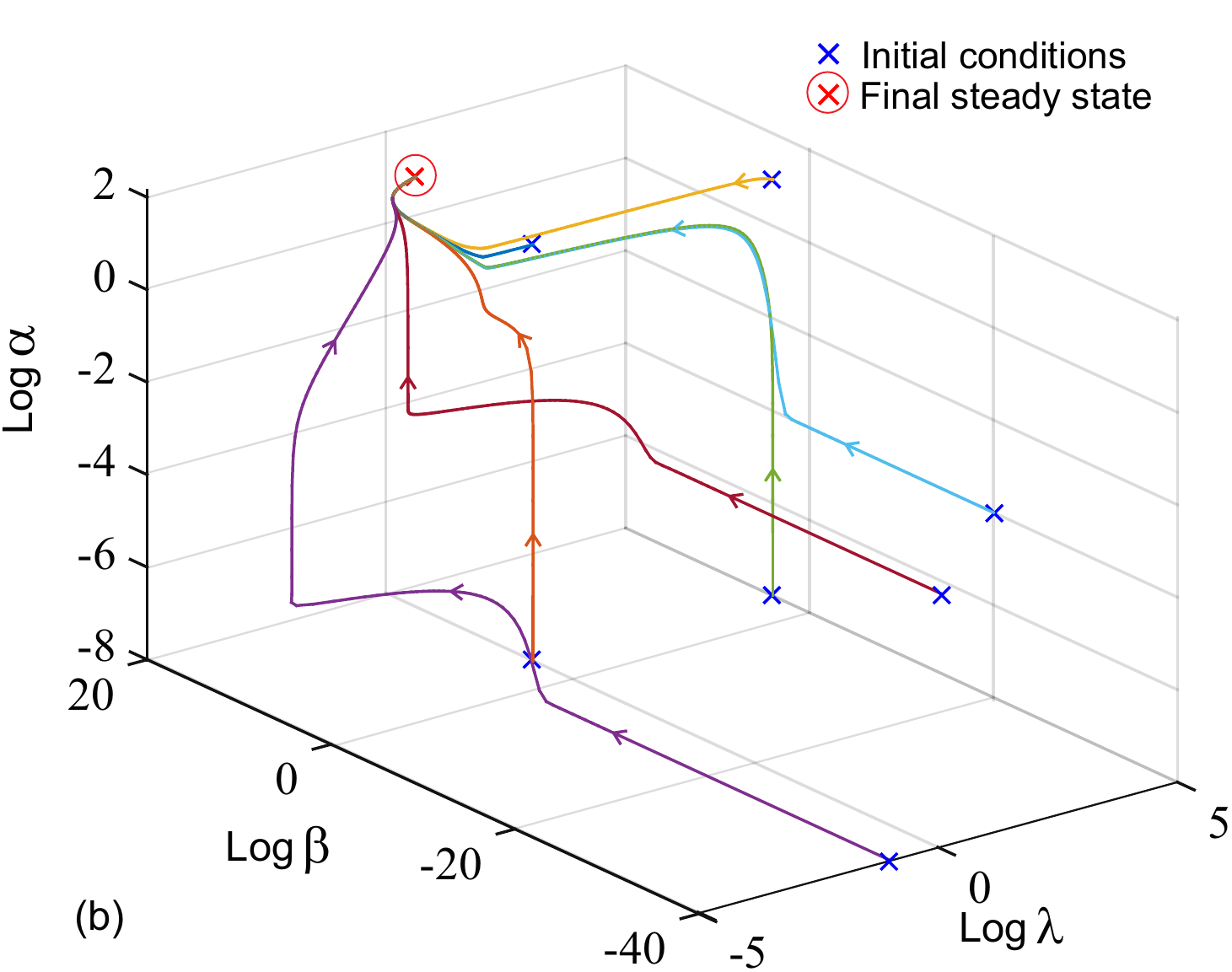}}
\caption{Graphs showing the evolution of $\alpha$, $\beta$, and $\lambda$ on logarithmic scales for: (a) low value of substrate stiffness $\tilde{\kappa} = 10^{-6}$, and (b) high substrate stiffness $\tilde{\kappa} = 0.1$,  from different initial conditions. The final steady state reached for (a): $\lambda=412$, $\beta <1$ and $\alpha=1.6$, and for (b):  $\lambda <1$, $\beta=108$ and $\alpha=2718$. \label{plot:Variable Evolution Map (3D)}}
\end{figure}

Figure \ref{plot:Time steady state reached} shows the dependence of the time taken for $\beta$ to reach steady state (relaxation time) on $\tilde{\kappa}$. On the whole, on soft substrates the relaxation time varies from about $10^{4.5}\rm\,s$ to a little over $10^5\rm\,s$ ($8$ to $30$ hours).  Experiments have shown that cells react to TGF-$\beta$ on time scales as long as $48$ hours (upper bound \cite{AssoianKomoriyaMeyersMillerSporn1983,PetrovFagardLijnen2002}) -- agreeing well with our results. It is interesting to observe how the time taken to reach the equilibrium steady state at high $\tilde{\kappa}$ increases substantially. At maximum possible values of substrate stiffness ($\tilde{\kappa} > 100$) the average value is about $10^{5.5}\rm\,s$ (or 3-4 days). It is possible that this high values of $\tilde{\kappa}$ are not practically achievable at all since rigid materials with a Young modulus $E \approx$ 1-10\,GPa have, in fact, a much smaller internal mesh size $\xi$. However, the overall trend of increasing the time to reach the homeostatic equilibrium with the substrate stiffness is a definite prediction of our theory.

\begin{figure}%[h]
\centering{
\includegraphics[width=0.9\columnwidth]{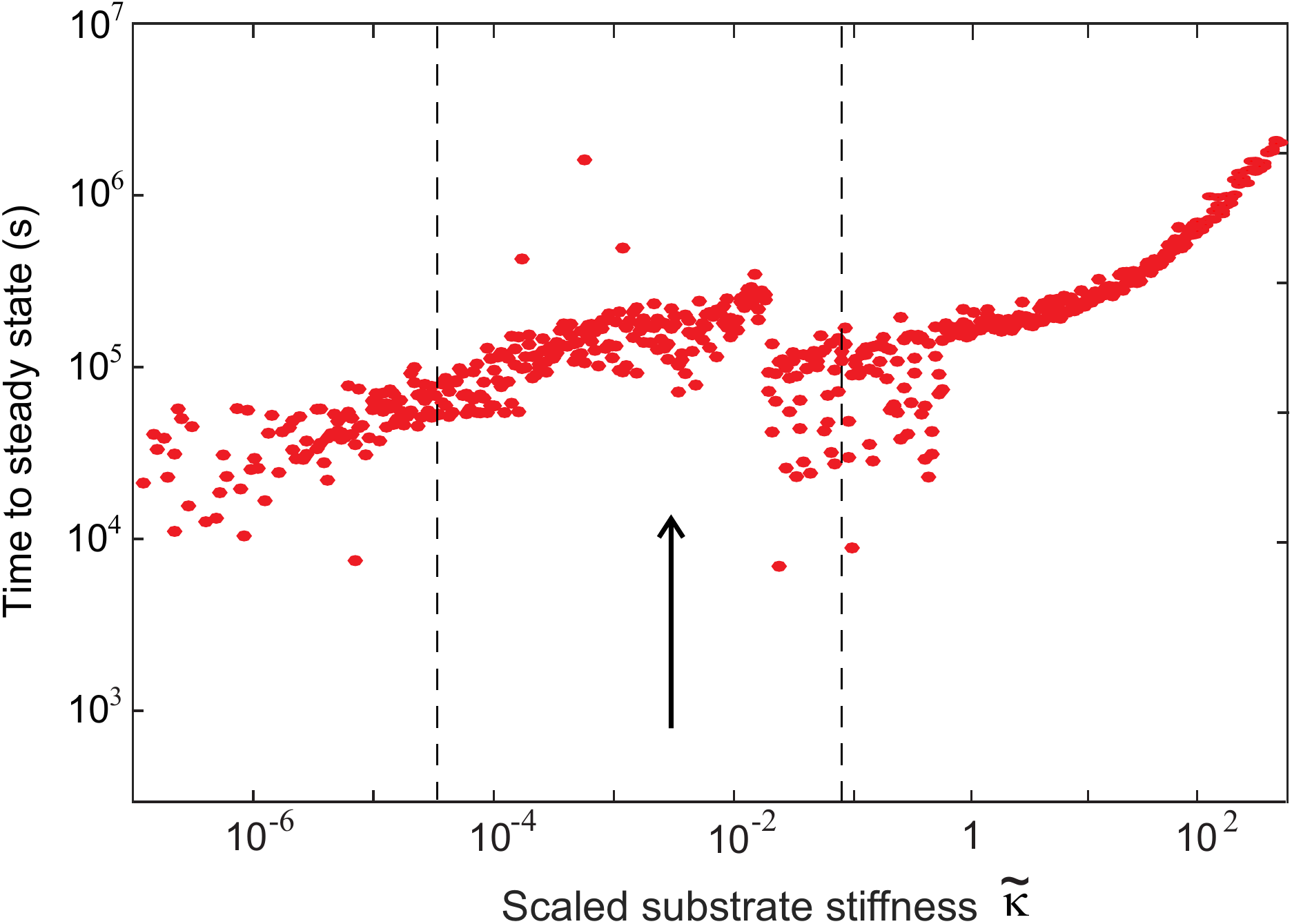}}
\caption{A graph of the time for each simulation to reach the equilibrium steady state, defined as when $\beta (t)$ reaches its equilibrium value within a tolerance of $1$ per cell, plotted against the effective substrate stiffness $\tilde{\kappa}$. Each point represents an independent simulation with randomly selected initial conditions, varied by four orders of magnitude. The arrow marks the experimental position of the myofibroblast transition and the dashed lines mark the `crossover region' of  Fig. \ref {plot:Variables_Vs_K}.
\label{plot:Time steady state reached}}
\end{figure}

\vspace{-10pt}
\section{The behaviour in the steady state}

The steady state dependence of each variable on $\tilde{\kappa}$ is of the highest importance in comparing our model to real biological systems and seeing how well it can describe reality. In a real cell, we would expect the concentration of TGF-$\beta$ to determine the response of the cell. Different concentrations would result in different gene expression or different enzyme activation and overall a different behaviour by the cell. We use the fibroblast system as the basis for discussion in this section.

Figure \ref{plot:Variables_Vs_K} shows how the concentration of each variable behaves in the steady state. There are three distinct regions: the low and high $\tilde{\kappa}$ regions where all variables are unchanging, and a middle $\tilde{\kappa}$ region which is the transition region.

\begin{figure}
\centering{
\includegraphics[width=0.85\columnwidth]{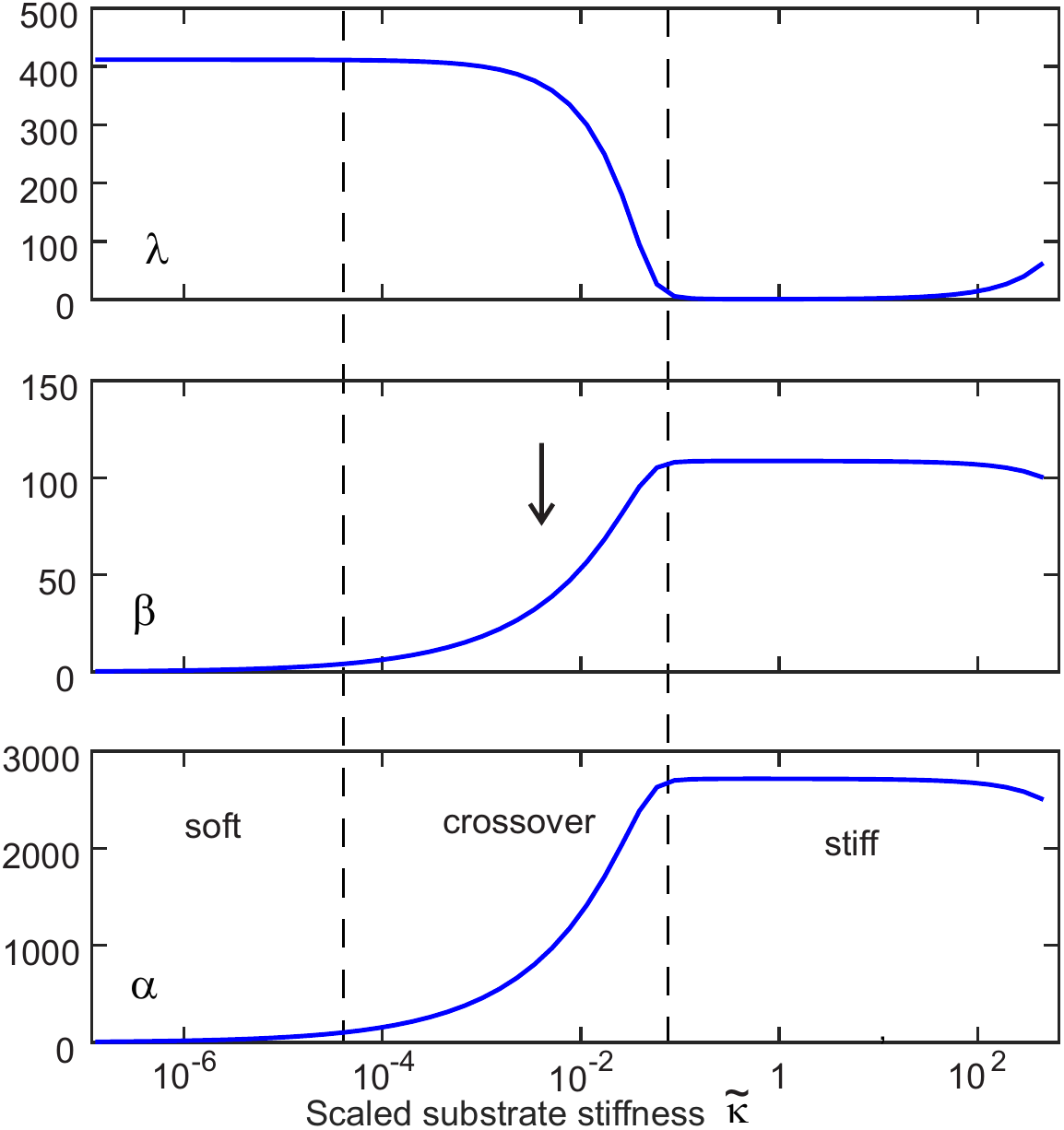}}
\caption{A graph showing variation of concentration of steady-state values of $\alpha$, $\beta$ and $\lambda$ (in units of particle per cell) against thescaled substrate stiffness $\tilde{\kappa}$ (expressed in decimal logarithmic units). Initial conditions were random. The dashed lines show the approximate location of the transition region and the arrow marks the position of myofibroblast transition.\label{plot:Variables_Vs_K}}
\end{figure}

\begin{figure}
\centering{
\includegraphics[width=\columnwidth]{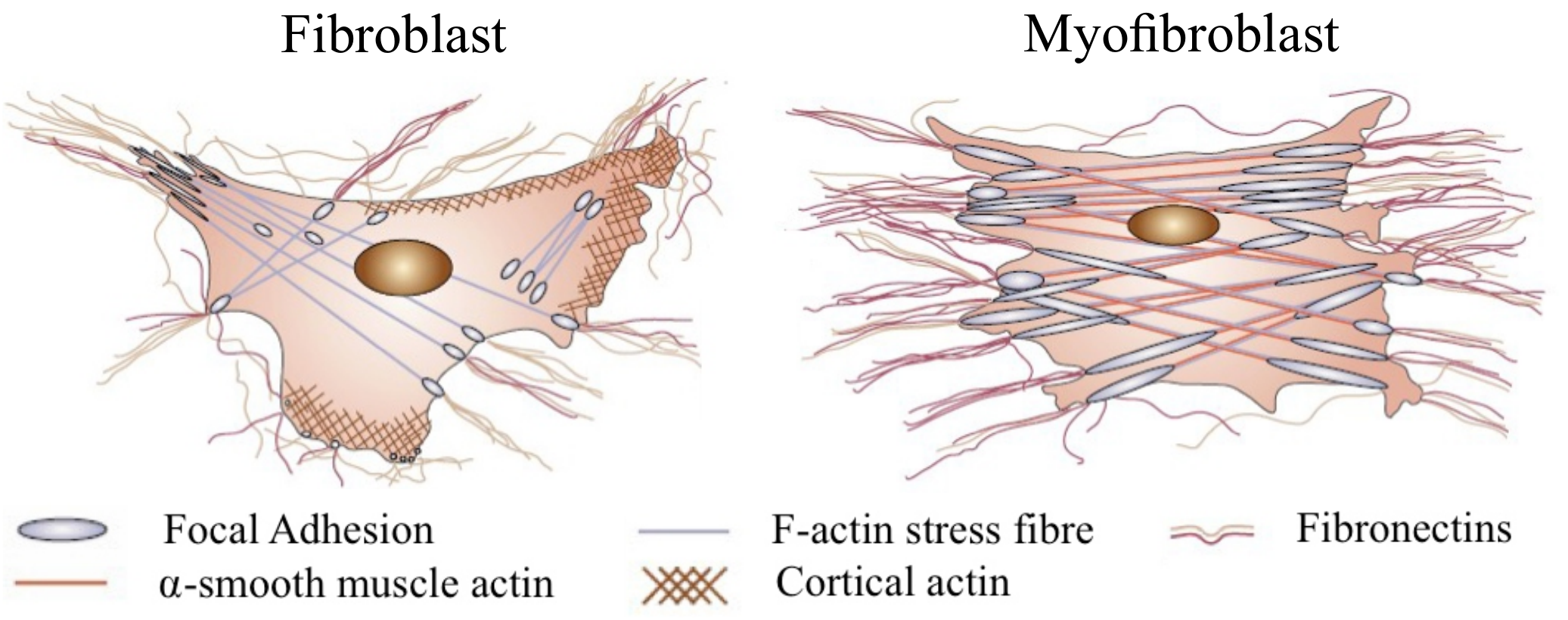}
%\rule{2cm}{2cm} % added by Karol Kozioł
}
\caption{A diagram of a contracted myofibroblast and less contracted fibroblast. Adapted from a discussion in  \cite{TomasekGabbianiHinzChaponnierBrown2002}. The large increse in $\alpha$-SMA expression is evident (as is widely reported increase in active TGF-$\beta$ on stiff substrates) -- but our conclusion about $\lambda$ (linked to the other two markers) suggests that there are only large focal adhesions, and no/few individual lc-TGF-$\beta$ on the cell surface on stiff substrates. \label{diagram:focal adhesion}}
\end{figure}

We first consider some peculiarities of the behaviour of $\lambda$. In the high $\tilde{\kappa}$ region we see the value of $\lambda$ tends to very close to zero, suggesting that as soon as lc-TGF-$\beta$ binds to the integrin and substrate -- it ruptures, so that on average there is none (or very little) present. This is a strange result and is certainly not intuitive. The reason for this is that in the high $\tilde{\kappa}$ region focal adhesions form (illustrated in figure \ref{diagram:focal adhesion}) where many lc-TGF-$\beta$ cooperate together in a combined structure. This is not accounted for in our model because we assumed that $\lambda$ represented only lc-TGF-$\beta$ attached individually to the substrate and integrin. Therefore we can deduce that only a small fraction of total lc-TGF-$\beta$ exist as individual elements in the high $\tilde{\kappa}$ limit -- a deduction supported by experiment \cite{han2012} showing that focal adhesion size increases on stiffer substrates. The true number of lc-TGF-$\beta$ may be much higher than that implied by the value of $\lambda$.

Examining the behaviour of $\alpha$ and $\beta$ in steady state, we see that the transition occurs in the region $10^{-4}<\tilde{\kappa}<10^{-1.5}$. The corresponding Young's modulus where the fibroblast transition occurs lies in between these two extremes, as required by our fitting parameter.

Over this transition $\beta$ increases from $6\rm\, cell^{-1}$  to a final value of $108 \rm\, cell^{-1}$. These values correspond to real concentrations of about $0.48\rm\, ng/ml$ and $8.6\rm\, ng/ml$, respectively -- taking TGF-$\beta$ molecular mass as $25\rm\, kDa$ \cite{AssoianKomoriyaMeyersMillerSporn1983} and assuming the fibroblast is a spherical cell with a radius of $5\rm\,\mu m$ (a big approximation given the myofibroblast is strongly contracted). How does this compare to real systems? TGF-$\beta$ concentrations as low as $0.1\rm \,ng/ml$ were found to be sufficient to induce a transition in endothelial cells to myofibroblasts with a maximal response at $1\rm \,ng/ml$ \cite{DoernerZuraw2009}. Another article \cite{PetrovFagardLijnen2002} observed stimulation of the fibroblast to myofibroblast transition for concentrations of TGF-$\beta$ between $0.1\rm\, ng/ml$ and $15\rm\, ng/ml$.

\begin{figure}
\centering{
\includegraphics[width=\columnwidth]{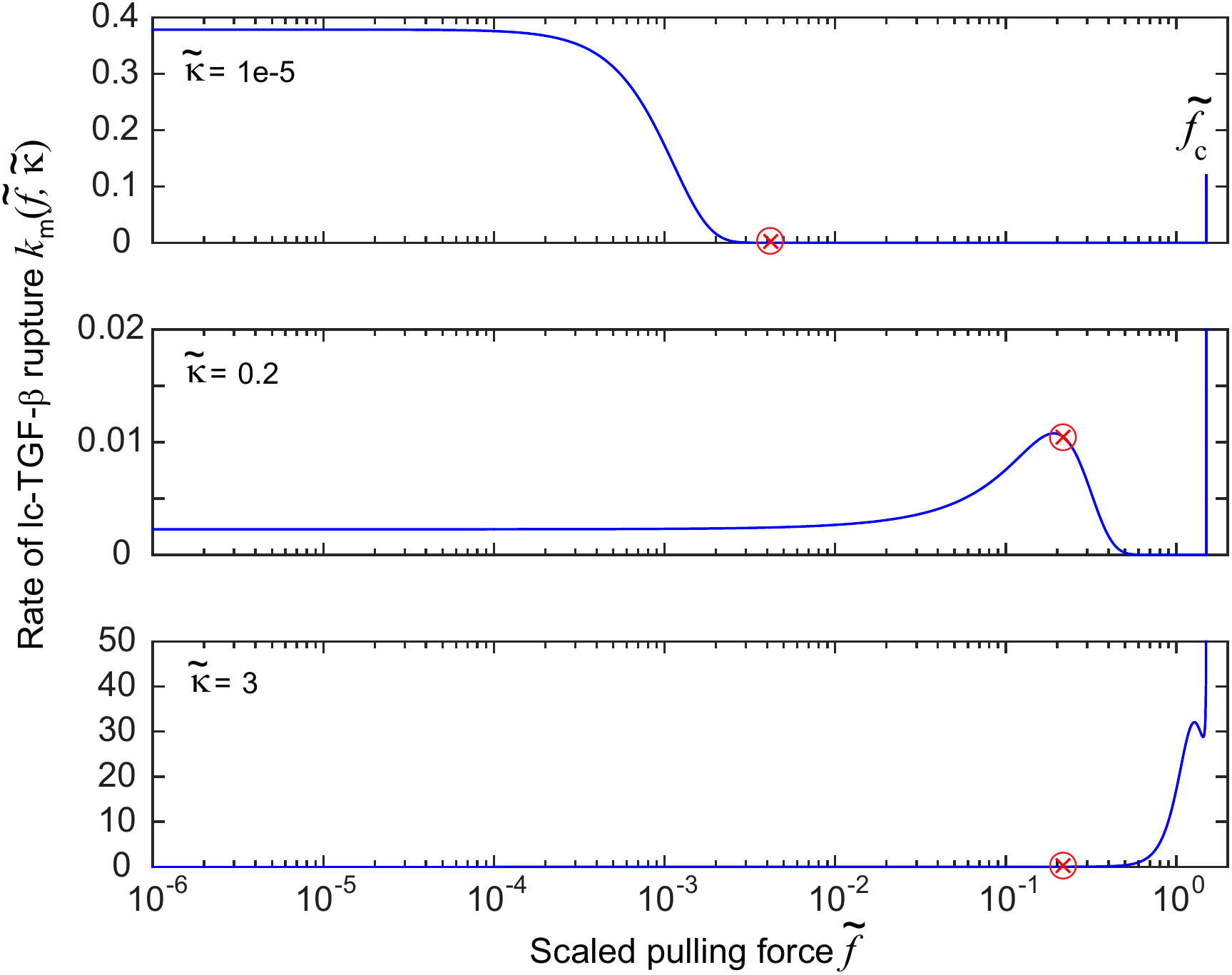}}
\caption{Plot of $k_{m}(\tilde{f},\tilde{\kappa})$ (in real units of [s${}^{-1}$])  for three different values of scaled substrate stiffness $\tilde{\kappa}$. The red cross represents the value of the dimensionless scaled force $\tilde{f}_\mathrm{eq}$ in cell equilibrium, corresponding to the steady state value of $\alpha$. In the lowest plot (for stiff substrate), the red cross corresponds to a value of $k_m=0.01 \, \mathrm{s}^{-1}$. \label{plot: Rate_Vs_f_AND_steadyState_alpha}}
\end{figure}

\begin{figure}
\centering{
\includegraphics[width=0.95\columnwidth]{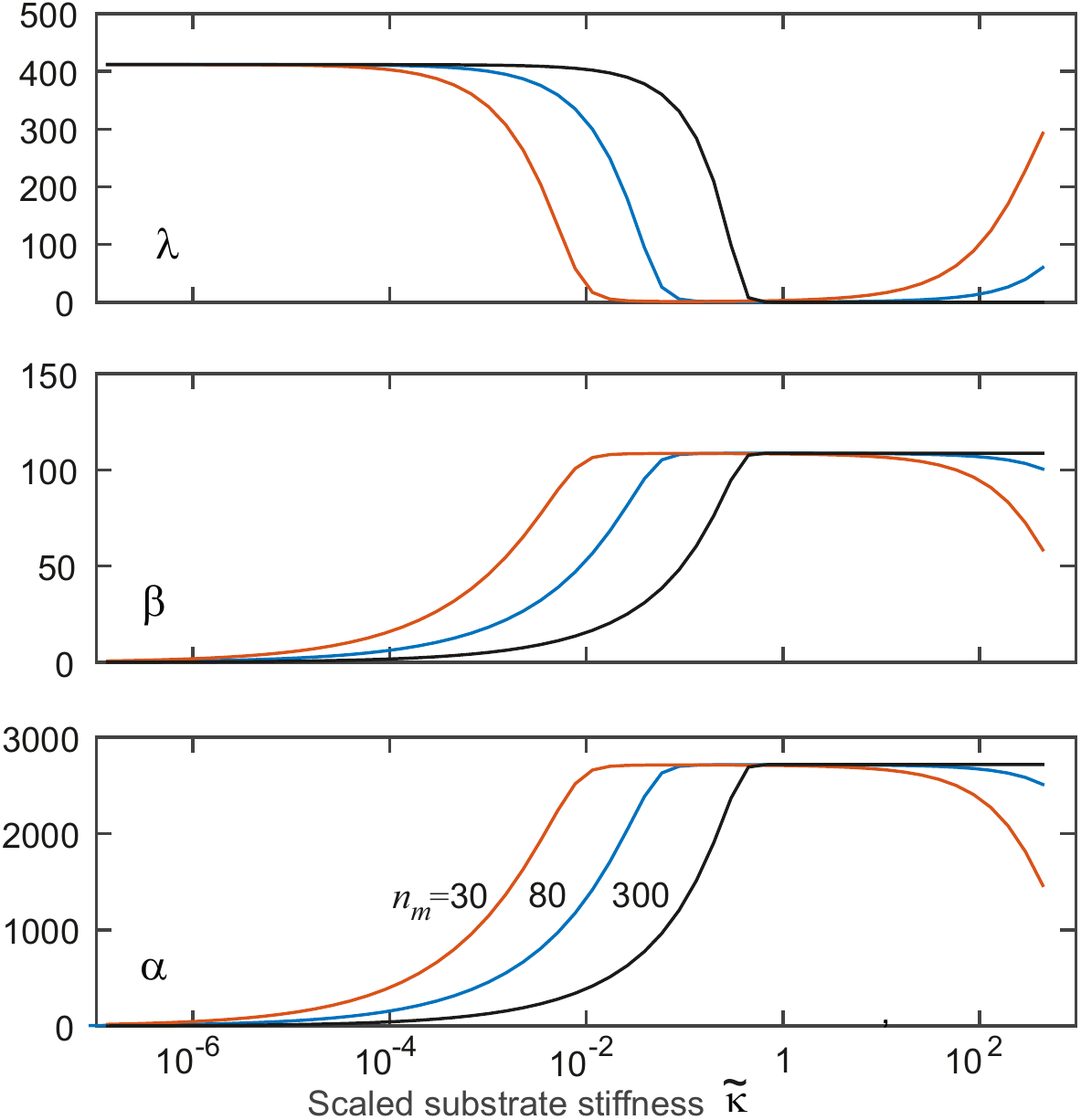}}
\caption{The effect of varying values of $n_m$ (the `number' of active myosin motors on the actin fiber) on the steady-state values of our three marker variables. The main effect is in adjusting the position of the transition region of high sensitivity to stiffness. The values are printed on the plot: $n_m=50$ (red), $n_m=80$ (blue), and $n_m=300$ (black). \label{plot:Changing_n_m}}
\end{figure}

\subsection*{The breaking rate $\bm{k_m(\tilde{f},\tilde{\kappa})}$ and applied force in equilibrium}

Finally, we examine the relation of force and rate of rupture of lc-TGF-$\beta$.  Qualitatively it would seem reasonable to expect the steady state to correspond to the maximum in $k_m$ (Fig. \ref{plot:Rate_Vs_f}) where the positive feedback changes to negative. Surprisingly we do not find this intuitive relation between system variables and $k_{m}$.

		Figure \ref{plot: Rate_Vs_f_AND_steadyState_alpha} shows the relation between the steady state value of $\alpha$ (directly proportional to force) and the rate of rupture of lc-TGF-$\beta$. As can be seen, system variables do not relax on the maximum in $k_{m}$. Instead there are three distinct types of behaviour. At very low $\tilde{\kappa}$ we see the steady state force is beyond the maximum in $k_{m}$ which is at $\tilde{f}\approx0$ (so the cell is not predicted to reduce its pulling to zero, as might have been assumed). At high $\tilde{\kappa}$ on rigid substrates, the steady state force occurs before the maximum, so in spite of its restructuring and formation of more stress fibers -- the coupled signalling loop stops the cell short of reaching the maximum rate of latent complex breaking. At intermediate values, in the `transition region' of $\tilde{\kappa}$, the steady state force rests somewhere after the peak of $k_{m}$. 

		We find that the transition in $\lambda$ from a high steady state concentration at low $\tilde{\kappa}$ to a low concentration at high $\tilde{\kappa}$ occurs between the low and intermediate cases discussed in this section. No particular effect in $\lambda$ is observed when the steady state value of $\alpha$ moves past the maximum in $k_m$. As far as the system has been tested, it seems that the maximum in $k_m$ has very little part to play in the dynamics of our coupled system variables.

		\subsection*{Using $\rm{n_m}$ to adjust the transition region.}

		The system proposed is clearly not sensitive to the substrate stiffness in the low or high $\tilde{\kappa}$ regions. However, it has been observed that cells respond to elasticities out of the range our model would suggest is possible \cite{englesSen2006}. We found that changing $n_m$ does not affect the steady state values of any variables away from the transition region. But it does affect the position of the transition region versus $\tilde{\kappa}$ -- as seen in figure \ref{plot:Changing_n_m}. If we allow the cell the flexibility to vary $n_m$ as well as the applied force, then $n_m$ could be adjusted so that the transition region scans over many orders of magnitude of $\tilde{\kappa}$. Myosin motor dynamics can be regulated by de/phosphorylation in-vivo \cite{watanabe2007}. Therefore, as long as the cell has a method to record $n_m$ and one of the variables, then it could dynamically vary $n_m$ to find the elasticity of its substrate.

\section{Conclusion}

		This work was a natural continuation of \cite{escude_rigozzi_terentjev2014} on the rate of rupture of lc-TGF-$\beta$. Incorporating their model into a full dynamically coupled system allowed us to create a set of differential equations with the mechanical sensing system encoded in this rate. The behaviour of the equation is entirely dependent on the applied force and the elasticity of the substrate, $\tilde{\kappa}$, and we have checked by starting from a large variety of random initial conditions that the equilibrium (steady state, homeostasis) of the cell is unique and specific on each substrate. We found that the system always returns to the steady state independently of initial conditions (a single stable attractor) and there appears to be a certain characteristic time scale of relaxation towards this equilibrium, regardless what the cell parameters have started with, of the order of $10^{4.5}$s (i.e. 8-9 hours), which is a broad agreement with in-vitro observations of cell deposited on various substrates.

		The choice of model parameters may appear the most controversial aspect of our work (since there are so many of them and some are little-known), yet we have taken great care to fix most of these parameters at specific values found in experiment. It is clear that for any particular cell system (osteoblast, fibroblast, smooth muscle, glial -- or indeed stem or various progenitors) the values of these parameters would be subtly different, and the cell response range vary accordingly. But the range of this variation is not great, and we are assured that the qualitative nature of the predicted response will remain the same. 

To summarise the predictions of our theory: 

1. \  We study a response of a `model cell' when it is placed on a specific substrate, with given elastic modulus and loss factor, or in other words -- stress relaxation time. The principal sensor is the large latent complex of TGF-$\beta$, which we use as a representative model. Following \cite{escude_rigozzi_terentjev2014}, we found the compact expression for the rate of latent complex rupture, $k_m(F,\kappa)$ in eq. \eqref{eq:Kmech0}, which is determined by the  parameters of the substrate and the steady pulling force from the cell. This breaking rate has a complex non-monotonic behavior with changing parameters, and it is central to the whole process of mechanosensitivity of the II${}^\mathrm{nd}$ kind. 

2.  \ On release of the signal molecule (active TGF-$\beta$) after the irreversible breaking of the latent complex, it could bind to a specific receptor on the cell to initiate the feedback loop -- or it could drift away from the cell. We have not yet  followed an obvious conjecture, that these `lost' TGF-$\beta$ signal molecules may provide an element of quorum sensing in a confluent tissue or cell colony. The concentration of free TGF-$\beta$ released from the broken latent complexes (our variable $\beta$) has specific values in homeostatic equilibrium. 

3. \  On placing the cell from any previous environment onto the given substrate (which we modelled by scanning a broad range of initial conditions), the cell remodels itself until it achieves the `equilibrium' morphology specific for this substrate. The most obvious and experimentally tested signature of this is the concentration $\beta$ in the vicinity of the cell, which is low/vanishing on very soft substrates, and high constant on mechanically rigid substrates (which is what has been registered in many experiments on very different cells and environments). The transition region in between these two limits is where the cell sensitively responds to small changes in stiffness by altering the equilibrium values of its parameters. 

4. \ The adjustment of cell morphology to its substrate stiffness happens via the expression of additional $\alpha$-SMA (stimulated by the TGF-$\beta$ binding), which is then incorporated into additional F-actin and stress fibers -- and provides more strong pulling force $F(\alpha)$ on the latent complexes in points of adhesion. Again, in close agreement with observations, we find that actin production is low on soft substrates and high on rigid substrates, with a transition region in between. The fibrosis in tissues is therefore linked to the stiffness of the environment the fibroblasts or smooth muscle cells are in contact with. 

5. \ Our final kinetic variable, the concentration of intact large TGF-$\beta$ latent complexes ($\lambda$) shows an interesting an unexpected response to stiffness change. On soft substrates, with low $\alpha$ (so the cell does not pull too hard on its adhesion points) and low $\beta$ (reflecting a low probability of lc-TGF-$\beta$ breaking) -- there are a lot of adhesion points around the cell: the equilibrium level of $\lambda$ is high. However, on rigid substrates when the cell pulls with a large force, we find $\lambda$ dropping to a very low value. Note that since the level of $\beta$ is high in this region, the breaking of latent complexes occurs very frequently -- almost immediately after they have been replenished by the cell, so there are almost no active adhesion points in operation around the cell surface.  This may sound contradictory, but we interpret it as the  commonly observed state when just a few large focal adhesion complexes hold the cell on a rigid substrate -- and we have now discovered that the cell surface outside of these few focal adhesions is free from the `singular' adhesion points. This is in contrast with the same cell on a soft substrate when it has a homogeneous concentration of singular adhesion points (high-$\lambda$ state). Our theory does not describe the collective effects that take place in large focal adhesions: certainly both the pulling force and the latent complex rupturing are very different there.

We have tried to make sure the stiffness transition region where steady state concentrations of $\alpha$ and $\beta$ rise by many factors, and $\lambda$ falls significantly, corresponds to observations for fibroblasts. Since most of the parameters we use are independently fixed, the critical role is played by the binding rate of TGF-$\beta$ to its receptor $k_{\beta}$, which we assume is not known. The value of  $k_{\beta} \approx 5 \cdot 10^{-5}\mathrm{s}^{-1}$ was required to bring this transition region to the Young modulus $ E \sim$ 10-15\,kPa  required to observe fibroblast to myofibroblast transition. However, we have shown that the cells have several mechanisms to adjust this transition region even for exactly the same lc-TGF-$\beta$ adhesion parameters: for instance by varying the number and the efficiency of myosin motors (the parameter $n_m$, Fig. \ref{plot:Changing_n_m}), or by channeling a different amount of newly produced $\alpha$-SMA to the adhesion pulling filaments (the parameters $\zeta$ and $k_\alpha$). Therefore we would expect that glial or neuron cells would `move' their stiffness transition to a much lower Young modulus range, while osteoblasts may benefit from their region of primary sensitivity to lie at a higher Young modulus as appropriate to their target environment. Perhaps this choice is made at the stem level and is at the core of subsequent differentiation:  by dynamic variation of $n_m$ by de/phosphorylation the cell can `scan' the transition region over many orders of magnitude of elasticity. 

		We found that the system had no particular dependence on the prominent and characteristic peak of the lc-TGF-$\beta$ rupture rate $k_m(F, \kappa)$, defying our intuitive expectation for how the system might respond. It seems that the general pattern of behaviour in homeostatic equilibrium is independent of many parameters -- at least to the extent that it was tested. 
Further work may seek to answer some of many questions posed by this paper. Does it matter that the peak in $k_m$ seems to not have the importance we expected? How could we include the effect of focal adhesions on lc-TGF-$\beta$ and would it change the behaviour of the model dramatically?

\subsection*{Acknowledgments.}  \ 
This work was supported by the EPSRC Critical Mass grant for Theoretical Condensed Matter,  the Sims Scholarship, and the Cambridge Trusts, and the University of Sydney. We are grateful to Drs. Boris Hinz, Sanjay Sinha and Kristian Franze for important discussions of their experimental systems. \\

%\bibliographystyle{rsc}
%\bibliographystyle{apsrev4-1}
%\bibliographystyle{elsarticle-harv}
%\bibliography{CellMechanosensing}


\end{document}